# Superlubric Brownian Motor

Keren Stein, Gautham Vijayan, Ron Bessler and Elad Koren*

*Faculty of Materials Science and Engineering, Technion – Israel Institute of Technology, 3200003 Haifa, Israel.*

*\*Email: eladk@technion.ac.il*

**Abstract -** Brownian motors are nanoscale machines that utilize asymmetric physical interactions to generate directed motion in space. The operation mechanism relies on the random motion of nanoscale elements generated by thermal activation. On the other hand, structural superlubricity (SSL) refers to a state of nearly vanishing friction due to structural mismatch between sliding interfaces. Van-der-Waals layered materials, such as graphene are of particular interest in this regard as they exhibit atomically flat surfaces and weak interlayer interaction. In particular, the sliding barrier in these systems turned out to be extremely sensitive to temperature, leading to the observation of thermal lubrication at elevated temperatures. Herein, the unique combination of a carefully designed tilted periodic potential landscape and virtually zero friction in incommensurate 2D layered systems are used to realize a mesoscopic superlubric Brownian machine. In particular, we perform mechanical shearing of superlubric graphite contacts to examine the influence of velocity on friction and adhesion. Our results show that while friction is virtually independent of velocity below 2500 nm·sec$^{-1}$, the conservative adhesion force increases by ~ 10 % with respect to the lowest measured velocity. Intriguingly, a greater amount of energy can be collected by the system once the retraction velocity is set above the protraction velocity. Our numerical calculations based on force field modelling indicate that a slow adiabatic sliding allows to utilize the available thermal energy to reduce the adhesion in agreement with our experimental observations. As a result, we demonstrate a mesoscale Brownian motor that can harvest thermal energy by adjusting the forward and backward sliding velocities and can pave the way for macroscale directed motion and energy harvesting.

**Introduction** - Brownian motors utilize thermal energy and asymmetric physical interactions to generate directed motion in space.[1,2] Analogue to molecular motors found in nature[3] (e.g. muscles, ions pump etc.), several artificial systems were able to utilize random thermal fluctuations to achieve directed motion in space based on optical tweezers[4], organic chemistry[5] and nanopatterning.[6] In general, the directed motion is motivated by an asymmetric or tilted[7] periodic potential landscape amplifying the probability to move in specific directions. Thus, although the available thermal energy allows the system to explore any possible excited state with equal probability, the presence of asymmetric landscape at non-equilibrium conditions results in a distinct bias towards a particular direction. Structural superlubricity (SSL)[8–14] refers to a state of nearly vanishing friction due to structural mismatch between sliding interfaces, with the potential of considerably reducing energy loss and wear.[15–17] According to theory, SSL is characterized by vanishing sliding force for infinite incommensurate flat surfaces, where the work done by one atom is fully compensated by the energy gain of another.[8–10] Consequently, the total friction force can completely vanish for adiabatic conditions that presumably take place at low sliding velocities.[10] 2-dimensional (2D) van-der-Waals (vdW) layered materials, such as graphene, $h$-BN and $MoS_2$, are of particular interest in this regard as they exhibit atomically flat surfaces and weak interlayer interaction.[11–14] In the last decade, many studies have considered the influence of different parameters affecting SSL including the dependence on interlayer orientation,[18–20] sliding direction,[21] applied load,[19,22] electric field,[23] contact size,[24–26] temperature[27] and velocity.[19,28–30]

The experimental relation between friction and velocity has been often realized by an atomic force microscope (AFM) tip sliding across 2D samples, providing an experimental assessment of a single asperity system.[30–36] The friction-velocity relation in few-layer graphene has been shown to depend strongly on thickness,[35,36] attributed to puckering,[36–38] or to its interaction with the substrate.[35,36,39] In addition, most studies observe a logarithmic dependence of friction with sliding velocity,[30–36,40] which is consistent with the thermally activated Prandtl-Tomlinson (TAPT) model.[31,41] According to the TAPT model, thermal activation adds energy to the system, which consequently assisting in overcoming potential barriers during sliding thereby reducing friction. Thus, slower sliding increases the influence of thermal effects, leading to lower friction. However, less focus has been afforded to the relation between friction and velocity under SSL conditions, in particular due to the intricate sample and experimental conditions that are needed to observe it. Experiments considering the basal plane of graphite,[29] graphite-DLC (diamond-like carbon)[28] and graphite-$h$BN contacts,[19] mostly show logarithmic dependence of the friction-velocity relations, whereas virtually zero dependence was observed for graphite and graphite-*h*BN contacts at velocities below 1 µm·sec$^{-1}$, which can presumably be attributed to adiabatic conditions.[10,40] The interlayer adhesion energy (binding energy) of several 2D materials including their heterostructures have also garnered significant attention,[24,42–45] due to their impact on nanoscale actuation[24], device fabrication, stability and performance.[46,47] Similar to friction, the direct measurement of the interlayer adhesive forces of 2D vdW layered systems turned out to be extremely challenging.[48] In contrast with friction, adhesion is traditionally considered velocity independent, yet experimental and theoretical work on this matter are scarce.[46,49,50] Importantly, the intriguing observation of self-retraction phenomena in graphite demonstrate the ability to almost fully conserve the total energy of the system while repeatedly breaking and repairing the interface.[48,51]

In this work, we utilize the unique combination of a carefully designed tilted periodic potential landscape and virtually zero friction in incommensurate sheared 2D layered systems to realize a superlubric Brownian motor. We start by studying the velocity dependent friction and adhesion of mesoscale 2D graphitic contacts in the superlubricity regime using an AFM setup, similar to our previous work.[20,23,24,52,53] We show that the friction dependence can be divided into two velocity regimes, where friction is virtually independent of velocity below 2500 nm·sec$^{-1}$ (Regime I), with a following logarithmic increase above this range (Regime II), similar to recent reports.[29,40] Moreover, the increase in friction is solely observed for the onwards direction (exfoliation), whereas no increase in friction is observed for the backwards direction (self-retraction), even up to the highest tested sliding velocity of 125000 nm·sec$^{-1}$. More intriguingly, by analyzing the adhesion forces at the low velocity regime (Regime I), we observe a continuous increase of up to ~ 10 % in the measured adhesion, that does not correspond to an increase in the dissipated energy throughout the exfoliation/retraction processes. Consequently, a substantial amount of energy can be gained by the system once the retraction velocity is set above the protraction velocity. Our numerical calculations based on force field modelling indicate that a slow adiabatic sliding allows to utilize the available thermal energy to reduce the adhesion, in agreement with our experimental observations. As a result, we demonstrate a mesoscale Brownian motor that can harvest thermal energy by adjusting the forward and backward sliding velocities.

The operation principle of the superlubric Brownian motor is schematically illustrated in Figure 1. The system comprises an actuated incommensurate bilayer graphene interface, where a simple mechanical spring (representing the AFM tip) is used for shearing the top section with a velocity $\bar{v}$ to reduce the interfacial overlap (Fig. 1A). The energy of the AFM spring as a function of distance can be described by a parabola i.e. $U = k \cdot (x-x_0)^2$ (solid black line, Fig. 1A), where $k$ and $x-x_0$ are the cantilever lateral stiffness and distance away from equilibrium, respectively. The crystalline atomic structure at the interface corresponds to a tilted periodic potential energy (solid red line, Fig. 1A), where the linear trend is attributed to the adhesive forces acting to increase the energy as the area overlap reduces. The corresponding asymmetric potential and the probability to move forward $P_F$ and backward $P_B$ during the exfoliation and retraction processes are presented in Figure 1B and C, respectively. In essence, in both cases there is a statistical preference for moving towards the right direction to release the mechanical tension of the spring. Hence, a larger probability is established for moving forward and backward during the exfoliation and retraction, respectively. The effective potential landscape is somehow analogue to the one presented in Figure 1D, which is a familiar element in Brownian machines.[1,6] In particular, it results in an asymmetric probability of a particle to move forward and backward across an energy barrier $E_b$ at $T > 0$ K i.e. $T \sim E_b/k_B$ (thermal energy approaches the energy of the sliding barrier), where $k_B$ and $E_b$ are the Bolzman constant and energy barrier height, respectively.

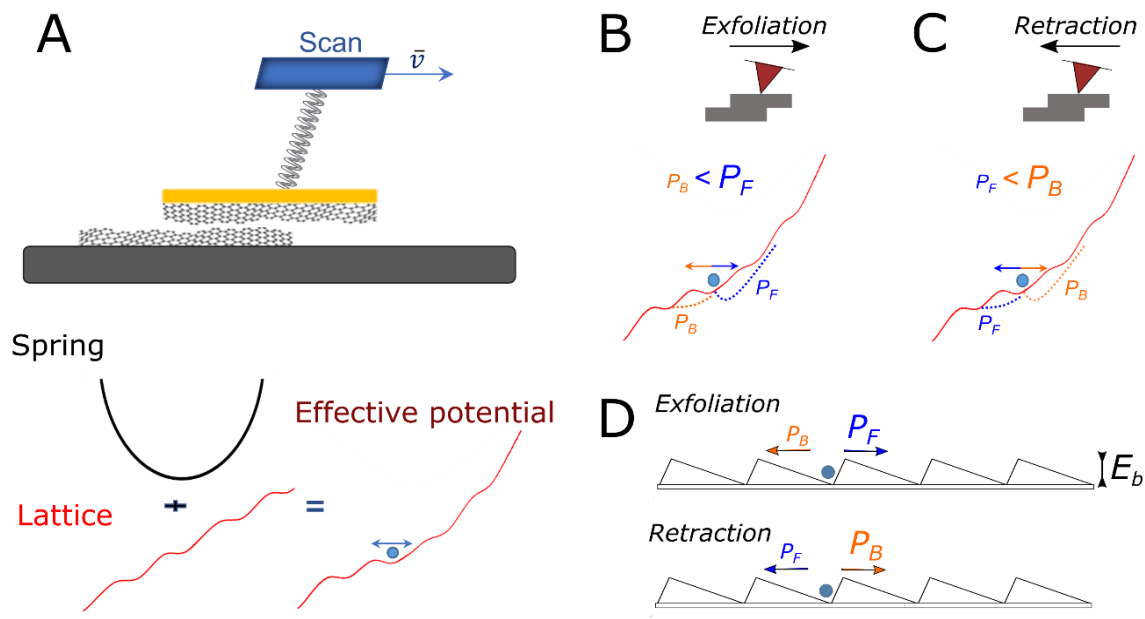

*Figure 1:* (**A**) Schematics of the Brownian motor system (top). Potential energies for the spring (parabolic) and graphite lattice (periodic) vs. displacement and the combined effective asymmetric potential (bottom). The asymmetric effective potential distribution and the probability to move forward $P_F$ (**B**) and backward $P_B$ (**C**) during the exfoliation and retraction processes. (**D**) Schematics of an asymmetric potential landscape and the probability of a particle to move forward and backward across the energy barrier $E_b$ at T > 0 K i.e. T ~ $E_b/k_B$ (thermal energy approaches the energy of the sliding barrier).

**Experimental results** - The dependence of adhesion and friction on sliding velocity was studied by means of electromechanical manipulation of mesoscale graphitic contacts (Fig. 2A). Graphitic contacts featuring cylindrical structures with a typical height of $100 \pm 10$ nm and a diameter of $210 \pm 5$ nm were constructed from highly oriented pyrolytic graphite (HOPG) based on a recently presented fabrication method.[20,23,24,52,53] The applied lateral forces induce a shear glide along a single basal plane in the HOPG structure. The lateral force was recorded for sliding velocities of 25 to 125000 nm·sec$^{-1}$ along a total distance equal to the mesa radius i.e. 100 nm (Fig. 2B). The total shear force is composed of a reversible displacement force due to adhesion and a smaller irreversible friction force characterized by force hysteresis. The small magnitude of the friction and the small force fluctuations (< 20 nN) indicate that the sliding is done under superlubric conditions.[24] Figure 2C presents the extracted force versus distance profiles for different sliding velocities, where the zero distance indicates the fully overlapped position of the top and bottom mesa structures. Both initial and final cycles are executed at 25 nm·sec$^{-1}$ and are virtually identical in terms of the average force magnitude and fluctuations range, thereby eliminating the influence of possible artifacts related to drift or an induced structural damage to the mesa structure throughout the mechanical actuation. The average applied force, that is attributed to adhesion, monotonically increases with sliding velocity, and seems to converge along the extreme values of the slower velocity profiles. The lateral force profiles across the right plateau region are presented in the enlarged image in Figure 2D. Figure 2E presents the complete measured set of the lateral force profiles at different sliding velocities within the range of 25 to 125,000 nm·sec$^{-1}$. Top (bottom) panel represents the measured forces for the case where the tip is sheared within the right (left) side of the pillar (as indicated by the inset schematics). It is clearly evident that the average sliding force for both trace (exfoliation, black profiles) and re-trace (retraction, blue profiles) directions increase (in absolute units) with increasing speed. Moreover, beyond a lateral velocity of ~ 2500 nm·sec$^{-1}$, the trace/re-trace profiles begin to separate as an evidence for increase in dissipated energy. Intriguingly, while all trace (exfoliation) force profiles keep growing with increase in velocity, the re-trace (retraction) profiles saturate along the maximum force magnitude of the lower velocity regime (marked by the dashed green line). This indicates that the increase in friction at high sliding velocities is solely associated with the exfoliation process, whereas retraction introduces virtually zero energy dissipation up to the highest tested velocity i.e. 125000 nm·sec$^{-1}$. We attribute this behavior to the intrinsically high self-retraction speed i.e. ~ millimeter per second of sheared graphitic structures in the superlubricity regime.[54]

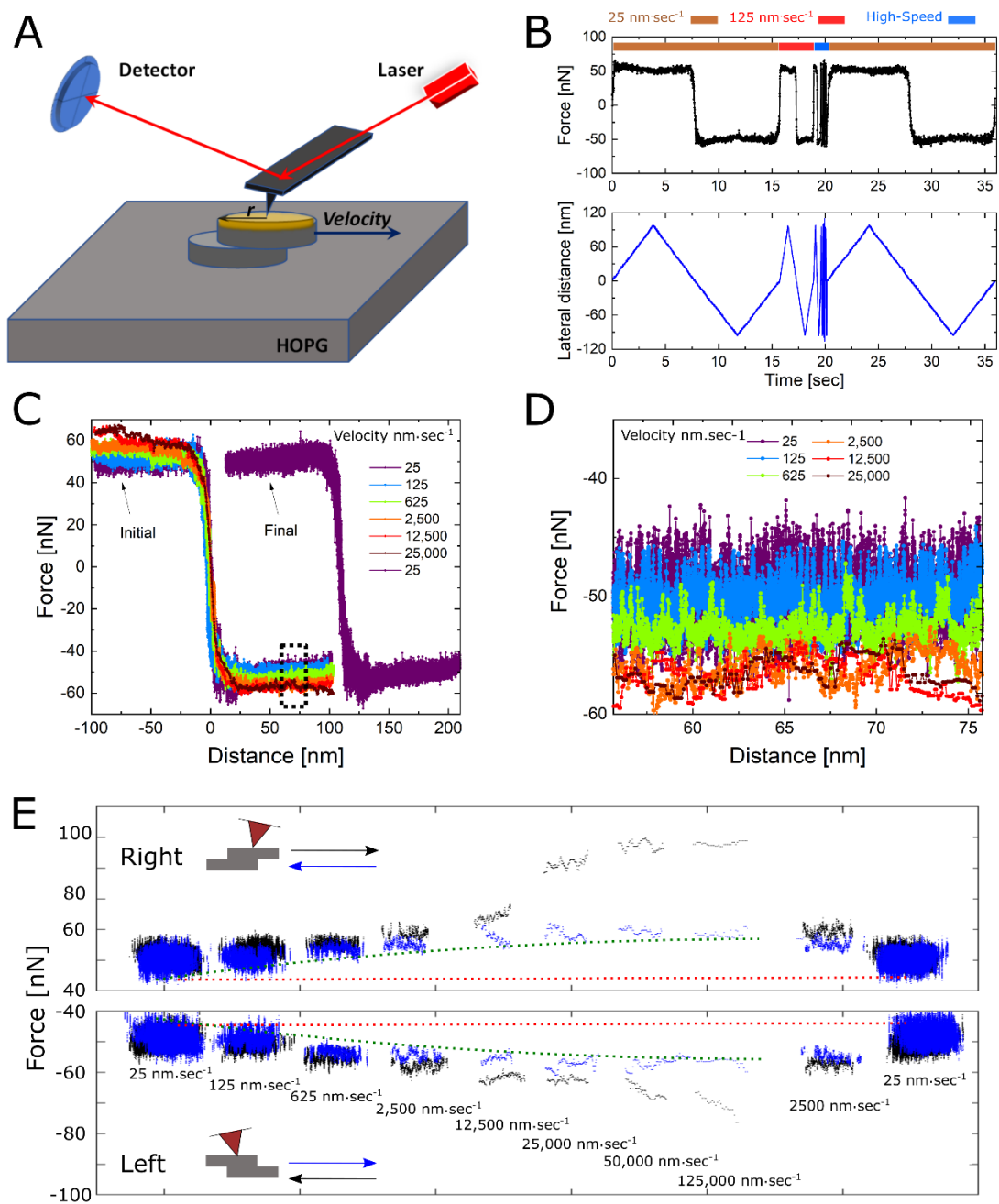

*Figure 2:* **(A)** Schematic description of the experimental setup. The AFM tip is cold-welded to the metal mask on top of the pillar, allowing to induce lateral shearing. The signal retrieved from the deflection of the laser beam represents the force applied by the tip at each point. **(B)** Measured lateral force (top) and distance (bottom) versus time while shearing the top part of the graphitic mesa structure at different velocities. Visible velocities displayed by the top color bar are 25 nm·sec$^{-1}$ (brown) and 125 nm·sec$^{-1}$ (red), while faster cycles of 625, 2500, 12500, 25000, 50000 and 125000 nm·sec$^{-1}$ (blue) are more difficult to observe in this timescale. **(C)** Force versus lateral distance for different velocities, taken from (B). To enhance visibility, the final cycle, which is conducted at a matching velocity to the initial cycle (25 nm·sec$^{−1}$), is offset to the right. **(D)** Enlarged section from panel (C) marked by the black panel, demonstrating that the force converges along the highest values with increased velocity. **(E)** Trace (black) and retrace (blue) force profiles measured at different sliding velocities. The profiles at the right side for 2500 and 25 nm·sec$^{-1}$ are repeated at the end of the experiment to confirm that the measured variations in force are not due to setup instabilities. Dashed red (horizontal) and green (non-uniform) lines are added as a guide to the eye to follow the adhesive force variations as a function of sliding velocities. Top (bottom) panel represents the measured forces for the case where the tip is sheared within the right (left) side of the pillar (as indicated by the inset schematics).

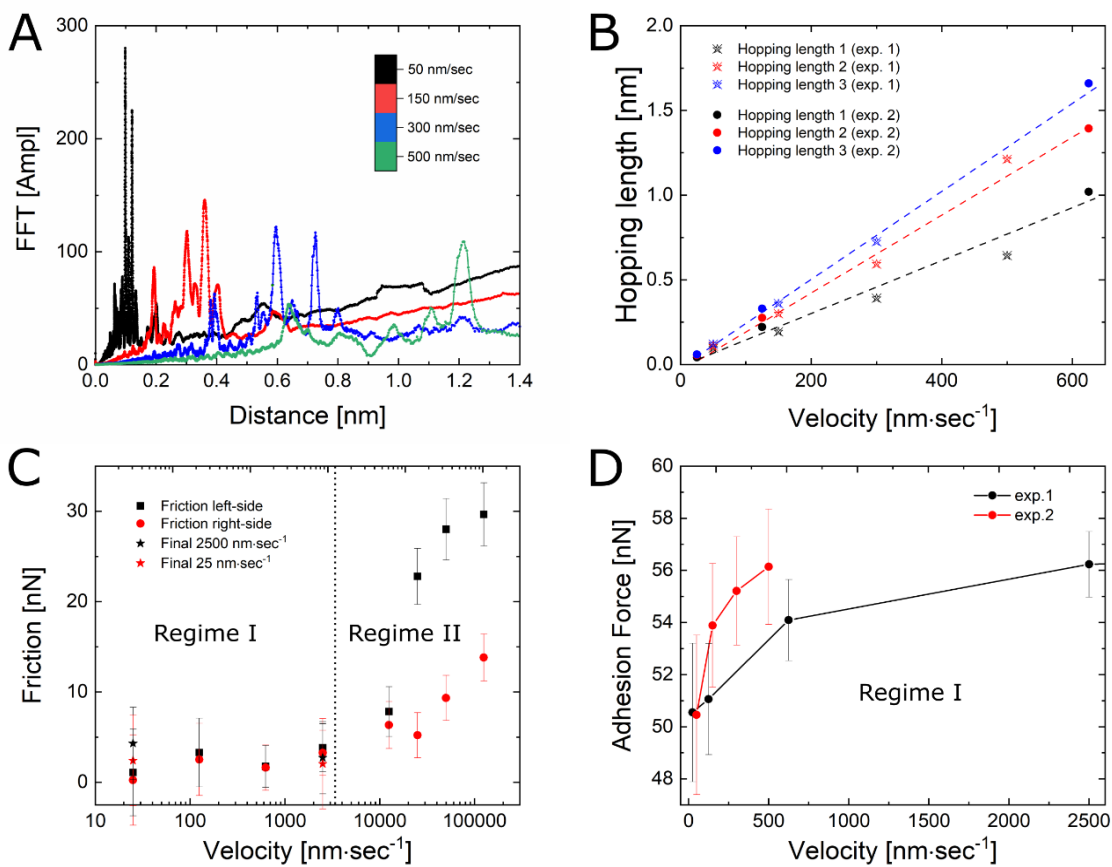

*Figure 3:* **(A)** Fast Fourier transformation of the measured force profiles for velocities: 50, 150, 300, 500 nm·sec$^{-1}$. Peaks represent typical hopping length between two consecutive energy minima. **(B)** Three most prominent hopping lengths per velocity. Data includes results from two different experiments i.e. exp. 1 and exp. 2. **(C)** Calculated average friction force for different velocities. "left-side" and "right-side" represent the direction to which the pillar is sheared before returning to the center. Friction is effectively constant below 2500 nm·sec$^{-1}$ (Regime I). The final sliding cycles for 25 and 2500 nm·sec$^{-1}$ are marked by star symbols (*). **(D)** Average adhesion force based on two experiments versus velocities related to Regime I (below 2500 nm·sec$^{-1}$), where friction is virtually zero and independent of the applied velocity.

Figure 3A shows the Fourier transformation of the lateral force versus distance, for velocities 50, 150, 300 and 500 nm·sec$^{-1}$. The observed peaks correspond to the typical hopping length between two consecutive energy minima and are in good agreement with previous report.[24] Interestingly, at higher velocities the profiles become more dispersed and are characterized by longer hopping distances. Figure 3B shows the correlation between the hopping length and sliding velocity for the three most prominent peaks of each velocity, where the distance between consecutive hopping lengths appears to grow linearly with sliding velocity. This suggests that at faster sliding velocities the contact overpassed more than one potential barriers per each hopping event. Figures 3C and 3D show the calculated friction and adhesion forces, respectively. The friction force was calculated separately for the displacement of the top mesa structure to the left (black symbols) and right (red symbols) sides with respect to the fully overlap position. The adhesion was considered as the difference between the average positive force (measured while the mesa was sheared to the left) and the average negative force (measured while the mesa was sheared to the right) divided by two. At velocities below 2500 nm·sec$^{-1}$ (Regime I), the friction force is virtually zero with respect to its standard deviation and with no measurable dependence on velocity, suggesting that the sliding is adiabatic. For velocities above 2500 nm·sec$^{-1}$ (Regime II), the friction force is characterized by a logarithmic growth, in accordance with previous research.[19,28–30,40] To adequately examine the velocity dependence of the adhesion force, we considered velocities below 2500 nm·sec$^{-1}$ (Regime I), in which friction is effectively independent of the sliding velocity (Fig. 3D). As demonstrated by the two separate experiments presented in Figure 3D, the average adhesion force grows by ~ 10 % with respect to the lowest measured velocity until it saturates. The saturation in adhesion follows the maxima force values measured for slower sliding velocities (Fig. 2D-E). The influence of velocity on the average adhesive forces is rather intriguing, since adhesion is typically considered independent of velocity based on thermodynamic arguments.[49,50,55] In fact, our experimental analysis indicates that a greater amount of energy can be collected by the system once the retraction velocity is set above the protraction velocity, which can facilitate a Brownian machine for energy harvesting.

In order to theoretically support our experimental results, we performed numerical simulations based on the force field theory of Kolmogorov-Crespi for graphitic interfaces.[56] The total potential $U_T$ includes both the lattice interaction energy $U_L$ and a simple mechanical spring $U_S$ that is used to shear the top graphitic contact, similar to the experimental method:

(1) $$U_T = U_L + U_S = U_L + k(x-x_0)^2$$

Where *k* is the spring constant representing the cantilever stiffness. In addition, the model includes possible relaxations in both the off-sliding axis Y and the mismatched angle θ due to preferred energetical orientations (see Methods section for more details). For simplicity, we assume rigidity of the surfaces and neglect surface relaxation. This assumption is rather rational under superlubric conditions, where the sliding can be approximated as adiabatic. In addition, out-of-plane distortions are significantly small for macroscopically supported contacts as opposed to single layer sliders.[35] The lattice potential energy is calculated by summing up the individual interactions between the carbon atoms within the contacted layers for each position throughout the slide. The driving force is the cantilever spring that moves at constant velocity. Hence, once the stored energy within the spring overcomes the energy barrier for slide, the equilibrium position $x_0$ will change. To allow for thermal excitations, we consider the Bolzman exponential occupation probability of higher energetic states. For each path option $(x, y, \theta)$ the energy difference $\Delta E_{xy\theta} = E_{xy\theta} - E_0$ between the next $(E_{xy\theta})$ and prior $(E_0)$ configurations is calculated and a path probability $P_{xy\theta} = \exp(-\Delta E_{xy\theta}/(k_B T))$ function is assigned to allow for thermal fluctuations. The actual path $(x, y, \theta)$ is chosen randomly according to the path probability weight. The model thus entails an intrinsic thermally activated randomness. Figure 4 presents the simulated results for two different velocities of the spring platform (Fig. 1A).

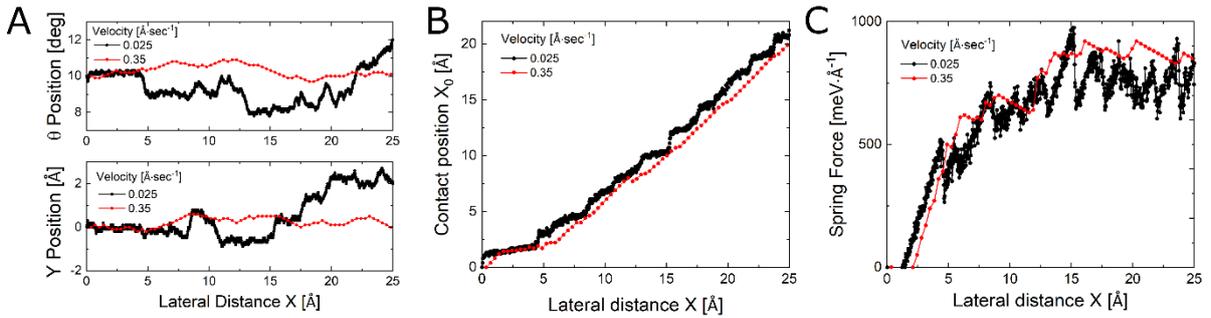

*Figure 4:* Simulated results for two different velocities conditions for 100 Å diameter bilayer graphene interface during shearing from the center position (full overlap). The initial misfit angle and the Y-position offset were set to 10 deg and zero, respectively. Temperature is set to 300 K. **(A)** Y-position and rotation misfit angle $\theta$ during the slide. **(B)** $x_0$-position of the top flake during the slide. At low velocity of 0.025 Å·sec$^{-1}$, the spring tends to jump forward earlier during the slide and therefore it is overall more relaxed compared to high velocity case of 0.35 Å·sec$^{-1}$. **(C)** Force versus distance profile, where the force profile of the higher velocity case follows the maximal values of the low velocity case, in agreement with the experimental results.

While both Y- and $\theta$ - positions present less features of the interfacial lattice interaction for the higher velocity of 0.35 Å·sec$^{-1}$ in comparison with the slower velocity of 0.025 Å·sec$^{-1}$ (Fig. 4A), the $x_0$ position for the higher velocity of 0.35 Å·sec$^{-1}$ shows larger lagging after the x position of the spring platform (Fig. 4B). Hence, we expect that the larger spring extension at higher velocity will correspond to higher spring force. Evidently, the simulated spring force versus lateral position is larger for the higher velocity of 0.35 Å·sec$^{-1}$ and follows the force maximal values of the slower velocity, in agreement with the experimental results (Fig. 2 C-D). Moreover, the force fluctuations for the higher velocity case often features hopping over two potential barriers, in good correspondence with the experimental results (Fig. 3A-B).

Similar results for initial angular mismatched configurations of 20 and 30 deg are included in the supplementary information. In order to better understand the origin of the velocity dependent characteristics, we consider next the effect of temperature over the sliding force characteristics. Figure 5 presents simulated shearing results at two different temperature conditions of 0 and 300 K. We present the off-slide axis Y- and misfit angle θ throughout the slide (Fig. 5A). Due to the higher temperature, the sliding at 300 K is less restricted to the minimum energy path. More intriguingly, the interfacial equilibrium position $x_0$ presents earlier movements forward for the higher temperature conditions, which indicates that the spring is effectively less stretched on average and thus consumes less energy during the slide (Fig. 5B). Figures 5C present the spring energy $U_S$ during the slide for the two different temperature conditions. The average spring energy (dashed horizontal lines in Fig. 5C) throughout the slide (average is considered between lateral distance of 15-50 Å) is slightly lower for the higher temperature case, indicating that thermal energy plays an important role in the reduction of the spring energy and consequently the effective adhesion energy of the graphitic interface. Figure 5D presents the total spring energy versus temperature for three initial angular mismatched configurations, in which a clear minimum in energy is evident for all three cases. Interestingly, the minimal energy point shifts to a higher temperature as the angular mismatch configuration reduces, presumably due to the relatively larger potential barrier of lower angular mismatched configurations. It is important to note that it is rather counterintuitive that at higher temperature the cantilever will be less stretched on average than at lower temperatures, as one would naively expect that the available thermal energy will facilitate the excitation of the spring to higher energy states i.e. larger tension. Indeed, at much higher temperatures thermal energy excites the spring to higher energy states leading to larger average tension and increasing energy consumption.

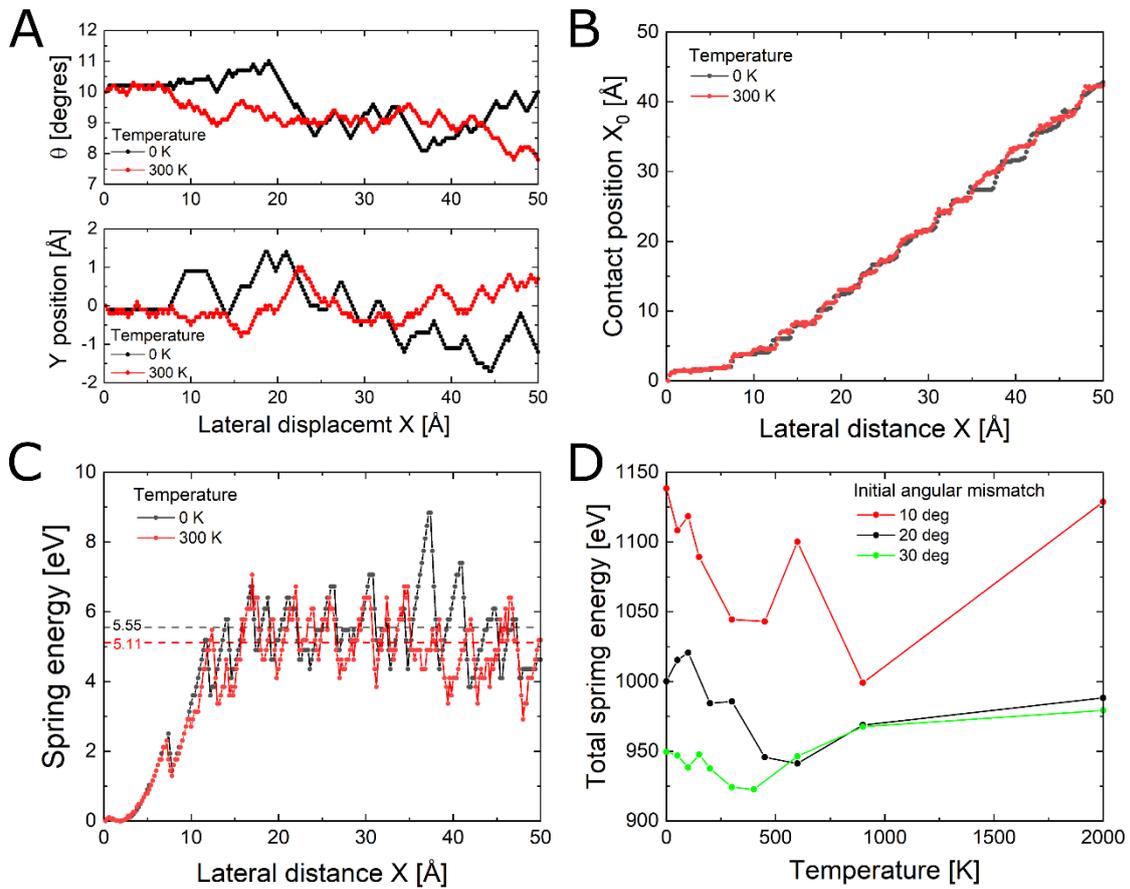

*Figure 5:* Simulated results for different temperature conditions for 100 Å diameter bilayer graphene interface during shearing from the center position (full overlap). The initial misfit angle and the Y-position offset were set to 10 deg and zero, respectively. **(A)** Y-position and rotation misfit angle θ during the slide. **(B)** $x_0$-position of the top flake during the slide. At 300 K, the spring tends to jump forward earlier during the slide and therefore it is overall more relaxed compared to 0 K. **(C)** Energy of the moving spring versus distance. Average values are marked by the dashed lines for each temperature (average is considered between 15-50 Å). **(D)** Total energy of the springs during the slide versus temperature for three different initial angular mismatched conditions.

## Methods

**Fabrication and shearing procedure -** The structures were fabricated from HOPG by means of reactive ion etching (RIE), using structured Pd–Au metal layers as self-aligned shadow masks. Nanomanipulation of individual nano-sized graphitic contacts was performed by an AFM in an N$_2$ filled glovebox (H$_2$O and O$_2$ content < 1 ppm). Cold-welding of a Pt/Ir metal-coated AFM tip to the metal on top of the mesa was established by applying a normal force and electrical current pulse of 50 nN and 1 mA, respectively, for 1 sec. The strong mechanical contact formed allows to apply lateral shear forces inducing a shear glide along a single basal plane in the HOPG structure. During the lateral manipulation of the graphitic mesas the normal force was kept below 5 nN.

**Lateral force constant calibration -** We use the adhesion energy of graphite $\sigma = 0.227 \, [\text{J} \cdot \text{m}^{-2}]$ to effectively calibrate the lateral force constant of the AFM cantilever.[24] Thus, we can directly use the graphite contact radii, $r$, to compute the measured lateral force *i.e.* $F = 2\sigma r$. The friction is calculated based on the difference in average force for trace and re-trace sliding directions. The adhesion is calculated as the difference between the average positive force (measured while the mesa was sheared to the left, including both trace and re-trace profiles) and the average negative force (measured while the mesa was sheared to the right, including both trace and re-trace profiles) divided by two.

**Simulations -** The potential energy is calculated according to the Kolmogorov-Crespi model,[56] which contains two components – short-ranged orbital overlap between the two layers causes exponentially decaying repulsion, while vdW attraction decays by $r^{-6}$:

$$V(\boldsymbol{r}_{ij}, \boldsymbol{n}_i, \boldsymbol{n}_j) = e^{-\lambda(r_{ij}-z_0)}[C + f(\rho_{ij}) + f(\rho_{ji})] - A\left(\frac{r_{ij}}{z_0}\right)^{-6}$$

$$\rho_{ij}^2 = r_{ij}^2 - (\boldsymbol{n}_i \boldsymbol{r}_{ij})^2$$

$$\rho_{ji}^2 = r_{ij}^2 - (\boldsymbol{n}_j \boldsymbol{r}_{ij})^2$$

$$f(\rho) = e^{-(\rho/\delta)^2} \sum C_{2n}(\rho/\delta)^{2n}$$

The hexagonal graphite lattice had an inter-atomic distance of a = 1.42 Å and lattice parameter of 2.46 Å. The circular graphite flakes had a diameter of 100 Å, and the initial non-commensurate rotation angle was set to 10, 20 and 30 degrees. The simulation begins from the fully overlap position of the bilayer graphene interface and then the top flake is pulled to the right by the spring for a lateral distance of 25 Å and 50 Å for the velocity and temperature dependent analysis, respectively. For each step the preferred interlayer configuration (account by the Bolzman distribution) was set by adjusting the $X_0$, Y and θ coordinates, based on 9 × X, 5 × Y and 5 × θ options relative to the previous conditions with a maximum change of ± 0.4 Å, ± 0.2 Å and ± 0.2 degrees, respectively (discretized steps of 0.1 Å along the X and Y axes and $0.1^0$ for rotation). The spring base velocity in the X - direction was kept constant i.e. 0.2 Å·sec$^{-1}$ for the temperature dependent analysis. The spring constant of the spring was set to 100 $[\text{meV} \cdot \text{Å}^{-2}]$.

**Acknowledgments –**

We gratefully acknowledge the Israel Science Foundation (ISF) grant 1567/18 for financial assistance and the Micro & Nano Fabrication Unit (MNFU) for the nanofabrication facilities. We thank Michael Urbakh, Oded Hod, and Astrid S. de Wijn for fruitful discussions.